%% file: main.tex
\documentclass[11pt]{article}
\usepackage[margin=1in]{geometry}
\newenvironment{IEEEkeywords}{\par\small\textbf{Keywords: }}{\par}
\usepackage{cite,url,array,graphicx,booktabs,tabularx,tikz}
\usetikzlibrary{positioning}

\usepackage[T1]{fontenc}
\usepackage{lmodern}
\usepackage{microtype}
\usepackage{longtable,pdflscape}
\usepackage[hidelinks]{hyperref}
\title{From Tool Connection to Execution Control: Benchmarking Security Invariants in MCP-Style Agent Runtimes}
\author{Ting Liu\\SymbolicLight Research\\\texttt{research@symboliclight.com}\\\url{https://github.com/SymbolicLight-AGI/handle-capability-protocol}}
\date{November 6, 2026}
\begin{document}
\maketitle
\begin{abstract}
Tool-connected agents need execution-time invariants that hold even when a model follows an adversarial instruction. We define eight invariants (\texttt{I1}--\texttt{I8}) for structured authority, canonical resource resolution, principal and delegation binding, handle lifecycle, explicit source-to-target transfer, request-correlated denial evidence, and single-use approvals. To test whether protection requires centralization or follows from these invariants, we compare HCP, a stateful handle-and-task runtime (\texttt{B2}), against an independent gateway ACL (\texttt{B3}) enforcing the same \texttt{I1}--\texttt{I8} contract. Both block all 18/18 modeled violations and complete all 25/25 benign workflows, with 16 ablations isolating each check. Three further studies show where execution control holds and why handle mediation matters. First, on 24 live AgentDojo tasks, DeepSeek reaches 4 attacker goals in direct runs while both controllers block 100\% of attacker calls; offline replay of 3,299 calls shows that 97\% of denied writes are substantive state or tool deviations rather than format noise (0/241 changed by format/case relaxation), and relaxing them admits 4 real exploits. Second, across 132 conditions on ChainCaps and Playwright MCP servers, per-call grants stop ungranted writes, and while session budgets stop plaintext leaks only by refusing unrelated work, explicit handle pipes and sealed reads block 100\% of handle leaks and post-revocation writes while preserving unrelated workflows. Third, on a matched MCP backend, HCP adds 0.26 and 0.31 ms to median file-copy latency in two passes (0.13 ms outside MCP calls, $>$25$\times$ faster than ChainCaps), trading 7.7 MB heap growth (10,003 audit records over 1,000 workflows) against 0.7 MB for \texttt{B3}. Identical decisions on shared benchmarks and zero reference-grant mismatches across 1,012 live calls (484 \texttt{B3}, 528 HCP) show that explicit \texttt{I1}--\texttt{I8} checks secure both designs.

\end{abstract}
\section{Introduction}
An agent can request an operation because a tool description says that it is useful, because a retrieved document contains a hidden instruction, or because a legitimate user explicitly asks for it. These explanations do not establish the authority to execute the operation. A useful security boundary must therefore examine the operation that would actually run, the authenticated caller, the resolved resource, and the applicable authorization facts.

Consider a user permitted to read a calendar note. An external description asks the controller to forward a secret to an attacker. A confirmation dialog may be accepted, yet the calendar grant still does not authorize the external destination. Conversely, a caller with a valid delegated grant may legitimately read a note owned by another user. A server that checks only ownership rejects that legitimate action. Both examples motivate examining execution decisions and utility together.

MCP standardizes interactions between clients and servers, including tools, transport behavior, and authorization guidance \cite{mcp_spec_2025_11_25,mcp_spec_2026_07_28,mcp_authorization_2025_11_25}. It does not prohibit server-side ACLs, reference monitors, or trusted gateways. Comparing a centralized runtime only against a weak connection layer (\texttt{B0}/\texttt{B1}) would conflate the value of execution-time invariants with a specific software topology. We therefore construct an independent gateway/server comparator (\texttt{B3}) that implements the full \texttt{I1}--\texttt{I8} contract—canonical resource resolution, delegated grants, handle ownership and lifetime, explicit transfer labels, and single-use approval binding—without sharing HCP's runtime or task machinery. Comparing HCP (\texttt{B2}) with \texttt{B3} tests architectural invariance across centralized and gateway/service deployments while isolating the state trade-off between HCP's handle-and-task mediation and \texttt{B3}'s inline checks.

We make four contributions. First, we state eight execution-layer security invariants (\texttt{I1}--\texttt{I8}) with explicit trusted components, controlled entrances, and counterexamples outside those entrances. Second, we provide a common observation and scoring procedure that separates unsafe effects, prevention, legitimate task completion, and request-correlated audit evidence. Third, we evaluate connection controls (\texttt{B0}, \texttt{B1}, \texttt{B1-decline}) and two independent \texttt{I1}--\texttt{I8} realizations (\texttt{B2} HCP and \texttt{B3} gateway/server ACL) using paired cases, actual source-copy ablations, and targeted mechanism witnesses. Fourth, we test how these invariants and their boundaries behave outside our scripted corpus: on held-out native AgentDojo tasks with live models (frozen payloads, an adaptive follow-up, and offline authority-rule replay); with unmodified third-party MCP servers in an isolated network; and in a matched cost study holding the real backend, inputs, and responses fixed.

The deterministic benchmark measures the consequences of scripted adversarial requests, not the probability that a model will generate them. The independent experiments add model-generated trajectories and real MCP services, exposing three boundaries: a live model may never attempt a given injection, strict reference-derived grants can constrain legitimate parameter variation, and per-call grants cannot inspect sensitive content that travels as an ordinary argument to an authorized destination.

\section{Background and Related Work}
\subsection{Connection protocols and execution authority}
The historical MCP 2025-11-25 path uses initialization before ordinary operations. The 2026-07-28 revision supplies protocol context per request and uses multi round-trip requests (MRTR) for supported server-to-client input requests. Its formal request metadata makes the protocol version and client capabilities mandatory; client information is optional descriptive metadata, not an authenticated identity \cite{mcp_meta_2026}. Our current test client supplies client information as well. A legitimate current request must not be rejected for omitting a historical initialization exchange.

The current MRTR specification defines an opaque string for request state and maps from request identifiers to input requests and responses. It mandates integrity protection when state influences authorization or business logic; principal, expiry, and original-request binding are recommended. Server-side single-use enforcement is mandatory when the relevant business operation requires at-most-once consumption \cite{mcp_mrtr_2026}. HCP implements a one-time approval operation and enforces these bindings. We do not claim that the protocol lacks them. Authorization-filtered capability discovery, in contrast, is an explicit deployment policy in this experiment. A public catalog is not intrinsically a protocol vulnerability.

\subsection{Agent attacks and enforcement systems}
InjecAgent studies indirect prompt injection through external content and evaluates tool-integrated model agents \cite{injecagent_2024}. MCPTox focuses on tool poisoning in MCP settings \cite{mcptox_2026}. These benchmarks motivate adversarial operations but answer a different question from ours: our controller deterministically issues a specified request so that execution checks can be measured independently of model behavior. We neither reuse their reported success rates as our baselines nor claim their workloads are interchangeable with this corpus.

Several recent systems already place controls at execution boundaries. ChainCaps propagates sink-specific capability budgets through explicit tool composition and limits its claims to trusted manifests and proxy-visible movement \cite{chaincaps_2026}. PORTICO, described in Lingering Authority, revokes resource-and-effect capabilities after trusted closure conditions and rejects stale uses before effects \cite{lingering_authority_2026}. CapSeal proposes a trusted broker that mediates secret use through constrained invocations and session binding \cite{capseal_2026}. AgentSpec provides customizable runtime enforcement for agent actions \cite{agentspec_2026}. These systems preclude claims that execution authorization, revocation, or mediation originate with HCP.

Information-flow defenses address a different part of the problem. CaMeL extracts control and data flow from the trusted user query so that retrieved data cannot change program flow, and it attaches capabilities to values to restrict exfiltration \cite{camel_2025}. Fides tracks confidentiality and integrity labels in the planner and enforces policies deterministically \cite{fides_2025}. AgentFlow specifies flow and path rules over labeled runtime edges, task-scoped capabilities, controlled release, and stateful taint, enforced by a runtime monitor \cite{agentflow_2026}. Per-call grant matching attempts none of this. AgentDojo supplies simulated email, travel, banking, and messaging environments whose task-specific checks judge utility and attacker success from the final environment and tool trace \cite{agentdojo_2024}. We use its unmodified tasks and checks as an external workload; we do not run CaMeL, Fides, or AgentFlow and do not compare against their reported numbers.

Our main comparator is not a reproduction of any of these systems. Their implementations, workloads, and trust assumptions differ. We implement a small independent ACL service and a gateway so that the same local operation sequence can be evaluated under comparable authorization facts. A separate follow-up calls ChainCaps's unmodified MIT-licensed core engine through a local bridge and composes it with that ACL; it does not reproduce the full ChainCaps proxy or published agent evaluation. We make no numerical performance or security comparison with a published workload that we did not run.

\subsection{Scope relative to information-flow control}
A label check at an explicit transfer entrance is weaker than tracking every value through an execution. HCP checks source labels when its data-pipe API receives a managed handle. By default, it does not propagate a label into all ordinary strings after an authorized full read. Thus, a later ordinary-argument call can satisfy destination authorization without carrying the source label. A separately evaluated trusted-source policy makes classified reads return pipe-only handles instead of plaintext. This closes the tested read-then-ordinary-argument route only while the caller lacks the plaintext through another path. The boundary experiments in Section 6 expose both behaviors. Preventing arbitrary movement of already known plaintext would require a stronger mediation or information-flow model.

\section{Security Model and Properties}
\subsection{Assets and adversary}
Protected assets are the contents of named notes and data handles, workspace write targets, simulated external deliveries, policy-scoped catalog entries, and acceptance of protocol operations requiring specific context. The adversary can supply natural-language descriptions and tool output, choose requested operations and arguments, present a handle obtained from another context, and copy, edit, delay, transfer, or replay an MRTR state string. In a header-consistency case, it can make the routing headers disagree with the JSON-RPC body.

The adversary does not choose the authenticated principal presented by the trusted host. A different-principal test uses a second correctly identified caller; it is not an identity-spoofing attack against an authentication system. Similarly, a supplied approval response represents the configured trusted host's decision. A valid MAC establishes state integrity and binding, not that a human genuinely reviewed or intended the action.

\subsection{Trusted components and controlled entrances}
The host, identity configuration, and approval channel are trusted. HCP's policy engine, grant store, data vault, task machinery, and in-memory audit mechanism belong to the trusted computing base. Local provider adapters are trusted to declare required scopes, effects, and accepted data categories accurately and to resolve the resource that their implementation will actually access. External natural-language descriptions and returned service content are not authorization facts.

A provider invocation receives a runtime object in the reference implementation. A malicious adapter executing in the same process could consequently invoke internal APIs or mutate shared state. HCP does not sandbox that adapter. Compromised in-process providers, forged host identities, dishonest resource resolvers, and malicious permission declarations are outside the evaluated adversary. Calling such an adapter merely ``partially trusted'' would obscure the assumption on which enforcement depends.

\begin{figure}[t]\centering
\begin{tikzpicture}[node distance=5mm,box/.style={draw,rounded corners,align=center,text width=.88\linewidth,minimum height=7mm,font=\footnotesize}]
\node[box,fill=gray!12] (external) {Untrusted descriptions, returned text,\newline requested actions and presented state};
\node[box,below=of external] (host) {Trusted host: authenticated principal\newline and approval channel};
\node[box,below=of host] (runtime) {Trusted runtime, grants, vault,\newline policy and in-memory audit};
\node[box,below=of runtime] (adapter) {Trusted local adapter: resolves actual resource,\newline declares scopes, effects and accepted labels};
\draw[->] (external)--(host);\draw[->] (host)--(runtime);\draw[->] (runtime)--(adapter);
\end{tikzpicture}
\caption{Trust boundary. External content cannot create grants. Same-process malicious adapters and forged host identities are outside the model.}\label{fig:trust-boundary}\end{figure}
The controlled entrances are capability invocation, authorized discovery when explicitly selected, handle access, explicit data-pipe transfer, and the implemented HTTP/MRTR protocol paths. We do not claim complete mediation of arbitrary JavaScript, operating-system I/O, or all routes through an external MCP server. The in-memory audit log is useful evaluation evidence, not durable or tamper-evident security storage.

\subsection{Properties and their assumptions}
\begin{table*}[t]
\centering\small
\caption{Properties, controlled entrances, and direct evidence. All require the trusted host and adapter assumptions.}\label{tab:invariants}
\begin{tabularx}{\linewidth}{lXX}\toprule
ID & Property and entrance & Direct evidence and limit \\ \midrule
I1 & Descriptions do not grant authority & Language-only perturbation; structured grants remain fixed. \\
I2 & Approval requires a matching grant & Approval without grant; granted write without approval. \\
I3 & Trusted canonical resource binding & Caller-hint conflict witness; adapter resolves actual resource. \\
I4 & Principal binding & Wrong caller, legitimate delegate, cross-principal state and handle. \\
I5 & Scoped capability invocation & Separate scope, capability, resource, expiry and revocation witnesses. \\
I6 & Explicit source/target transfer checks & Owner/lifetime and PII acceptance tests; ordinary arguments are outside label propagation. \\
I7 & Current-attempt rejection evidence & Audit removal and prior-call substitution; no durability claim. \\
I8 & Explicit protocol and approval context & Historical initialization, current metadata/headers and bound single-use state. \\
\bottomrule\end{tabularx}
\end{table*}
For an invocation with principal p, capability c, resolved resource r, and time t, the grant predicate requires an active grant whose principal, capability, scopes, and resource all match. Approval is an additional condition for configured effects; it cannot make a missing grant valid. Legitimate delegation is represented by a grant to the delegate for the owner's resource. Ownership alone is therefore not the authorization predicate.

For a full handle read, possession of the identifier is insufficient: the owner recorded by the trusted vault must match the caller, and the handle must remain live. Under the separately tested pipe-only release policy, even the authorized caller cannot directly display classified handle content. Metadata-only introspection is a distinct API path and is not covered by the full-content confidentiality claim. For an explicit pipe, the runtime checks source ownership and lifetime before destination authorization and the PII acceptance rule. It permits ordinary strings to reach an authorized destination without reconstructing their provenance; Section 6 treats that behavior as a limitation rather than silently extending the property.

These properties do not imply eight independent switchable components. For example, changing only natural-language text should leave an authorization decision unchanged because authorization uses structured facts. A property test demonstrates that independence. Removing the grant check changes several properties at once. We therefore distinguish the property being evaluated from the implementation check removed by an ablation.

\section{Reference Implementations}
\subsection{HCP execution and effect observation}
HCP represents capabilities, grants, managed handles, and tasks in a single in-memory runtime. Before invoking a trusted provider, it resolves the actual resource through the adapter and evaluates the applicable grant and effect policy. Caller-supplied resource hints do not override that resolution. Tasks expose completion, waiting, or failure, but these statuses are not our effect oracle.

The experimental providers use the same fixture world as the comparators. Reads return the fixture's actual content. Writes append the target and full content to a local write ledger; simulated deliveries append the destination and payload to an external-delivery ledger. Dry-run writes return a preview and append no actual write or delivery. These ledgers are test instrumentation: they model local effects without contacting a remote service or writing sensitive files. Their content lets the observer distinguish an incorrect target, an empty payload, a legitimate first write, and an additional replayed write.

\begin{figure}[t]\centering
\begin{tikzpicture}[node distance=4mm,box/.style={draw,align=center,text width=.88\linewidth,minimum height=6mm,font=\footnotesize}]
\node[box] (request) {Caller, capability, input and approval};
\node[box,below=of request] (resource) {Trusted adapter resolves actual resource};
\node[box,below=of resource] (grant) {Current grant match + effect approval};
\node[box,below=of grant] (effect) {Provider effect: observed content and target};
\draw[->] (request)--(resource);\draw[->] (resource)--(grant);\draw[->] (grant)--(effect);
\node[box,below=of effect,fill=gray!12] (pipe) {Explicit pipe additionally checks source owner,\newline lifetime and PII acceptance before invocation};
\node[box,below=of pipe] (limit) {Full read then ordinary arguments:\newline no automatic source-label propagation};
\end{tikzpicture}
\caption{Invocation and transfer checks. The ordinary-argument route still requires destination authority, but does not preserve the explicit pipe's source-label restriction.}\label{fig:execution-flow}\end{figure}
\subsection{Request-correlated audit}
Each invocation or protocol attempt receives an independent correlation identifier. Runtime events emitted during that attempt inherit the identifier through asynchronous request-local context. HTTP rejections before dispatch record the rejection stage, method when known, reason, and request correlation. If the authenticated identity is not yet available, the record explicitly reports that condition; a fixed placeholder string does not count as an authenticated subject.

MRTR verification failures produce a record for the failing attempt. This includes integrity, subject, request, expiry, replay, and malformed-response failures. The prior legitimate call's log cannot satisfy the later attempt's audit criterion. Existing event schemas remain available; the scorer translates their field names without inventing absent facts. Protocol rejection and resource authorization use different required field sets. State strings, authentication tokens, and protected content are not added to the new denial records.

\subsection{Limited MRTR approval path}
The implemented modern path covers an approval round trip for \texttt{tools/call}. The server returns an \texttt{elicitation/create} form requesting a Boolean approval, keyed by a server-issued input-request identifier, after checking that the client declared form-elicitation support; the client retries with a new JSON-RPC identifier, the unchanged opaque state, and a response map. The state carries integrity-protected principal, capability, request digest, input-request identifier, expiry, and nonce. A resumed request is validated without consuming the state; a missing input is requested again and a malformed response is an explicit error. A valid response consumes the nonce synchronously before any provider effect, only an explicit true approval proceeds, and current grants are rechecked, so revocation after issuance prevents execution. This gives at-most-once recovery within one process, not a distributed or crash-durable mechanism.

\subsection{Comparator configurations}
B0 is a connection-layer lower bound with no execution authorization. B1 enables the existing description lint, server ownership check, approval prompts, content scanner, protocol gating, header consistency, and connection logs; lint and scanning are advisory unless a sensitivity variant enables blocking. Its operator approves effect prompts. B1-decline changes only that approval policy. These are local configurations, not claims about every deployed MCP server.

B2 is HCP. B3 is an independently implemented gateway and service ACL that enforces the complete \texttt{I1}--\texttt{I8} invariant suite without importing HCP's policy or runtime code, omitting only HCP's centralized task scheduler and in-memory vault graph. The trusted gateway binds the caller and validates approval and protocol context (\texttt{I1}, \texttt{I2}, \texttt{I8}), while the service resolves the canonical resource (\texttt{I3}), matches caller, capability, scopes, delegated grants, expiry, and revocation against active ACL entries (\texttt{I4}, \texttt{I5}), verifies handle ownership and lifetime (\texttt{I5}), checks trusted source-to-target labels on explicit transfers (\texttt{I6}), and emits request-correlated rejection records (\texttt{I7}). Both B2 and B3 receive identical grants, labels, requested actions, and destination policies, testing whether \texttt{I1}--\texttt{I8} protection is architecturally invariant across a centralized runtime and a gateway/service split.

\section{Evaluation Method}
\subsection{Corpus and paired setup}
The fixed main corpus contains 18 attack cases and 25 legitimate workflows, retaining their historical identifiers. Cases cover adversarial metadata, confused delegation, approval without a grant, explicit data transfer, historical initialization, scoped discovery, handle misuse, header/body disagreement, and MRTR replay or transfer. The appendix lists each case, intended violation, group, and observed outcome; ancillary files provide machine-readable scores.

We distinguish execution and data-access properties, catalog deployment policy, and protocol or state-binding properties. The historical transport subset and the current request subset are configured independently. A case assigned to the current path does not imply that its underlying attack concept exists only in that version. Handles and discovery policies are useful across multiple protocols. MRTR state binding is grounded in the current formal specification, not an inference from an older proposal.

The controller, initial notes, destinations, grants, labels, and operation sequence are shared across modes. The controller attempts the specified adversarial action deterministically; metadata is not treated as proof that a model would obey it. Each case starts from a fresh world. Preparation actions are separated from the evaluated attempt. For a replay case, the observer records an authorized first execution and then checks whether the later attempt produces an additional effect.

Legitimate tasks include own-resource reads, delegated reads, approved writes and deliveries, internal transfers, dry-run previews, authorized discovery, an owner-handle read, historical and current protocol flows, and an MRTR approval round trip. Success requires the expected content and target, or the relevant protocol output. Multi-step tasks require their ordered component goals. Dry-run success additionally requires absence of real effects. The catalog utility test requires that authorized requested entries remain visible; disclosure of other entries is assessed separately by the policy attack cases.

\subsection{Shared outcome rules}
Implementations return responses, observable effects, and actual logs. A shared observer evaluates a goal predicate over that evidence. Attack success means that protected content reaches the tested caller or sink, a target write occurs, a policy-protected catalog entry is returned, or a disallowed protocol operation is accepted. Prevention requires absence of the protected effect and an explanatory control outcome, such as denial, withholding, or waiting for approval. An empty revoked-handle response can therefore prevent disclosure even without a denial event.

Audit sufficiency is scored independently. For resource or authorization decisions, a matching current-attempt record must identify the subject, capability or operation, resource or handle, decision, and reason. For an early transport rejection, identity may be unavailable, but the record must explicitly say so and include the method, reason, decision, and correlation. The scorer reports missing fields rather than deriving completeness from an HTTP status, task status, or error code.

An initialization failure, timeout, unexpected exception, or unmet preparation condition is inconclusive. It is not evidence that the attack was prevented. Legitimate noncompletion is categorized by observed cause, including a policy rejection, waiting for input, or an execution problem. We reserve false rejection for a legitimate task actually rejected by policy; we do not use it as a name for every incomplete workflow.

\subsection{Actual ablations and witnesses}
We evaluate 16 mechanism-removal variants. Each uses a disposable source copy and an explicit patch manifest with exact match counts and before/after file hashes. A patch that does not match exactly stops that variant. The unmodified copy must reproduce the workspace before the removal variants are interpreted. No public runtime option disables these checks, and the experiments never patch the working runtime in place.

Every variant runs the complete main attack and legitimate corpus. We report newly successful attacks, remaining prevention, legitimate-task changes, missing audit evidence, and inconclusive executions. A later check can still stop an attack after an earlier check is removed. That is a measured composition effect, not a failed experiment. Removing approval can also prevent an MRTR preparation phase from generating an approval challenge; those trials remain inconclusive rather than being forced into a predicted outcome.

Separate witnesses isolate checks that the main corpus does not distinguish. Examples include a correct handle owner with an expired handle, a fully authorized read with a conflicting caller resource hint, and a valid same-principal state whose request body changes. MRTR integrity, request binding, and expiry witnesses retain all other relevant conditions. They supplement rather than replace the full corpus. Natural-language perturbation is tested as a property, not modeled as a fictitious independent enforcement module.

\subsection{Reproducibility}
The locally retained results record their format version, run identifier, source and fixture hashes, environment, observations, evidence links, and missing audit fields. A validator recomputes scores and summaries and rejects missing modes, mismatched hashes, broken evidence links, and contradictory effects; fixture expectations are never edited to make a run pass. We repeat the deterministic matrix twice and compare responses, effect ledgers, and audit structure after normalizing timestamps and random identifiers. Historical results are kept separately and do not enter current denominators.

\subsection{Separate boundary, HTTP, and composed-baseline follow-ups}
Three follow-ups keep their own source-bound results. In the sealed-boundary experiment, trusted adapters return a pipe-only (\texttt{pipe\_only}) handle for a classified note, so a caller with a read grant cannot peek its plaintext. Paired HCP and ACL cases cover PII-rejecting and PII-accepting pipes, a handle or already known plaintext passed as an ordinary argument, a missing destination grant, an internal note, and a legitimate request to display the classified note. A local HTTP experiment timed the implemented MCP 2026-07-28 \texttt{tools/call} path \cite{mcp_spec_2026_07_28,mcp_transports_2026}, and earlier in-memory operation timings used 100 warm-up and 1000 measured operations per process; the appendix reports both timing summaries. The composed baseline keeps B3's checks and adds the unmodified MIT-licensed ChainCaps core engine through a local bridge, with trusted source manifests that assign downstream sink budgets to fixture labels. It reruns the fixed cases and tests ordinary-argument composition after a source read; it does not run ChainCaps's MCP proxy or its published model-driven tasks.

\subsection{Independent tasks, real services, and cost}
Three further experiments test whether the conclusions depend on our corpus, scripted adversary, and fixtures. Each has its own denominator, and we do not rank results across them.

The first uses native AgentDojo tasks with live models \cite{agentdojo_2024}. From the pinned public source, excluding composite tasks, a fixed hash order selected 8 development and 24 test tasks in 4 domains; their native utility and injection checks are applied unmodified to the final environment and the executed tool trace. The evaluated agents are gpt-6.1-sol and deepseek-v4-pro. Direct execution runs every call. The ACL (\texttt{B3}) and HCP (\texttt{B2}) receive identical authorization facts: every read tool, plus the mutating operations and target fields that the native reference solution performs in that condition's environment. These grants enforce checked reference fields rather than inferred intent; a field mismatch alone establishes neither an attacker operation nor a legitimate equivalent.

Payloads occupy only the benchmark's attacker-controlled fields. On the development tasks we compared 4 families: the benchmark's important-instructions and tool-knowledge attacks, a task-context continuation framing the attacker goal as a pending prerequisite, and a source-document substitution that replaces the attacker-controlled record, which we treat as source poisoning rather than instruction hijacking. A preregistered rule ranked them by direct-mode attack effects and selected source-document substitution and task-context continuation. Selection, protocol, corpus, and source and dependency hashes were frozen before the first test request, and the runner refuses to start if a frozen file changes.

Per session we record exposure (the payload reached model-visible output), attempt (\texttt{Att.}, a later call that, applied to an isolated copy of the environment, satisfies the native attack check), blocking of such a call (\texttt{Blk.}), the native attack goal (\texttt{Goal}) and user-task utility (\texttt{Util.}) checks on the final environment, and inconclusive endings (timeouts, truncation, turn limits), which never count as prevention; an effect before an inconclusive ending counts. Interception is defined only over attempts. The task is the unit: contrasts average matched sessions within a task, and per-task differences are summarized with a task bootstrap interval, a sign-flip test, and per-domain values. An independent validator re-executes each saved trace on the saved initial environment, verifies the final environment, reruns the native checks, and audits the write-ahead request ledger.

An adaptive follow-up reuses these controllers and grants on 24 fresh tasks, selected in a fixed hash order from tasks outside the first experiment's development and test sets, with three further models: gemini-3.6-flash-high, deepseek-flash, and GLM-5.3-flash. Round one places a deterministic task-context continuation, built from the goal and the reference calls, in the native injection vector. If round one is scored and misses the native attack check, a fixed generator (deepseek-flash) writes one revised payload from the round-one transcript for a second session. The protocol, payloads, and scorer were frozen before the first request, and tamper tests confirm that the independent validator detects planted edits. Exploratory offline replay of the recorded ACL and HCP trajectories from both native studies classifies each denied write against same-condition reference candidates into four equivalence classes—state-equivalent (\texttt{Eqv}), substantive difference (\texttt{Diff}), unresolved (\texttt{Unres}), and no same-tool reference (\texttt{NoRef})—and evaluates format/city/identity normalization (\texttt{V1}--\texttt{V3}) and an approve-all diagnostic control (\texttt{V4}) under both static pre-state re-evaluation and first-divergence trajectory replay.

The second experiment uses real MCP servers: the unmodified ChainCaps frontend and backend \cite{chaincaps_2026} and Microsoft Playwright MCP \cite{playwright_mcp}, in a separate network namespace with a loopback receiver. Twelve prescribed workflows cover secret copies to a file or HTTP sink, legitimate and private copies, unrelated actions after a sensitive read, missing and wrong labels, known plaintext, handle revocation, and an ungranted target, under direct execution, per-call ACL, HCP, an ACL with a session source budget matched to ChainCaps's policy, native ChainCaps, and HCP and ACL with explicit pipes or sealed reads. Outcomes are recomputed from workspace files and raw receiver lines.

The third experiment measures cost on the same backend with identical inputs and responses: a file copy and an HTTP relay under nine modes at one, four, and sixteen workflows in flight, with fifty warm-up and one thousand measured workflows per cell, in two independently shuffled passes. We separate backend launch, controller setup, backend time, and the remainder, and track controller state and heap over one thousand workflows.

\section{Results}
\subsection{Effects, utility, and evidence}
\begin{table*}[t]
\centering\small
\caption{Main results. Attack columns use the fixed attack denominator; benign columns use the legitimate-workflow denominator. Audit means sufficient current-attempt evidence, independently of effects. Uncertain trials are not prevention.}\label{tab:main-results}
\begin{tabularx}{\linewidth}{lrrrrrr}\toprule
Mode & Success & Prevented & Uncertain & Attack audit & Benign done & Benign audit \\ \midrule
B0 & 18/18 & 0/18 & 0/18 & 0/18 & 25/25 & 0/25 \\
B1 & 11/18 & 7/18 & 0/18 & 14/18 & 24/25 & 19/25 \\
B1-decline & 3/18 & 14/18 & 1/18 & 13/18 & 14/25 & 19/25 \\
B2 & 0/18 & 18/18 & 0/18 & 18/18 & 25/25 & 21/25 \\
B3 & 0/18 & 18/18 & 0/18 & 18/18 & 25/25 & 25/25 \\
\bottomrule\end{tabularx}
\end{table*}
B0 permits 18 attacks. B1 permits 11 and prevents 7. Its revoked-handle response prevents the protected content from being returned, although that attempt lacks sufficient audit. A score requiring a denial log would incorrectly convert this prevention into an unblocked attack. Deleting audit entries in the scorer-independence test leaves both effect classifications unchanged.

HCP (\texttt{B2}) and the gateway/server ACL (\texttt{B3}) each prevent 18 main-corpus violations and complete 25 legitimate tasks. Their agreement reflects a shared invariant contract rather than coincidence: because \texttt{B3} enforces the same \texttt{I1}--\texttt{I8} checks (including canonical resource binding, handle ownership/lifetime, explicit transfer label checks, and single-use approval state) at the gateway/service boundary, it achieves the same protection as HCP's centralized runtime. HCP's attack-attempt audit meets the declared criterion throughout this corpus, while audit coverage on legitimate tasks is 21 of 25 because ordinary successful public protocol flows do not emit the detailed authorization records required on rejection paths.

B1 completes 24 legitimate workflows. The rejected case is an authorized delegated read that its ownership-only check does not recognize. The repaired modern protocol setup and dry-run effect observation avoid attributing old harness errors to the controls. B1-decline completes 14 legitimate workflows and rejects 11 by policy. It permits 3 attacks, prevents 14, and has 1 inconclusive replay trial: the blanket decline policy prevents the legitimate first write needed to establish the consumed-state setup. We report that unmet precondition instead of labeling it successful replay defense.

The appendix groups these attacks into execution and data checks, the selected catalog policy, and protocol or state conditions; grouping avoids conflating deployment policy with protocol conformance. In particular, withheld catalogs are evaluated only under an explicitly selected scoped-catalog policy. The public-catalog witness returns the catalog and is considered correct. The HTTP header cases evaluate routing consistency \cite{mcp_transports_2026}. The MRTR cases evaluate the selected integrity and one-time approval contract and are not evidence that all implementations must expose exactly HCP's internal state design.

\subsection{Data-flow boundary}
\begin{table}[t]
\centering\small
\caption{Separate data-flow boundary experiment. Caller, source, target resource and payload are fixed. The accepting-target row changes only trusted acceptance metadata; the missing-grant row changes only target authority. These rows do not enter the main attack denominator.}\label{tab:boundary-results}
\begin{tabularx}{\linewidth}{Xrr}\toprule
Path / target policy & HCP write & ACL write \\ \midrule
pipe-reject-pii & No & No \\
pipe-accept-pii & Yes & Yes \\
peek-ordinary-parameter & Yes & Yes \\
ordinary-missing-grant & No & No \\
\bottomrule\end{tabularx}
\end{table}
With a fixed caller, source, and authorized target, both execution-authorizing implementations reject an explicit PII pipe when the target does not accept that category and permit the same data when the trusted target declaration accepts it. An authorized full read followed by ordinary-argument invocation writes the same content even when the explicit pipe would be denied. Removing the target grant still prevents that ordinary invocation.

These observations identify the scope of the source-label rule. They are not included in the protected-pipe attack success denominator. Ordinary-parameter transfer exposes this boundary; its code and raw observations remain unpublished. Neither grant matching nor approval converts this mechanism into full information-flow control.

\subsection{Mechanism removals}
\begin{table*}[t]
\centering\small
\caption{Actual isolated HCP source removals. Each variant executes the complete attack and legitimate corpus. New success is relative to the unchanged reference; uncertain trials are retained. Targeted witnesses are additional evidence.}\label{tab:ablation-results}
\begin{tabularx}{\linewidth}{Xrrrr}\toprule
Removed check & New success & Benign done & Uncertain & Witness changed \\ \midrule
unchanged & 0 & 25 & 0 & 0 \\
grant-matching & 6 & 25 & 0 & 10 \\
canonical-resource & 0 & 25 & 0 & 1 \\
approval & 0 & 24 & 3 & 1 \\
handle-principal & 2 & 25 & 0 & 1 \\
handle-lifecycle & 1 & 25 & 0 & 2 \\
datapipe-class & 1 & 25 & 0 & 0 \\
audit & 0 & 25 & 0 & 0 \\
legacy-initialization & 2 & 25 & 0 & 0 \\
protocol-version & 0 & 25 & 0 & 1 \\
catalog-filter & 2 & 25 & 0 & 0 \\
header-consistency & 2 & 25 & 0 & 0 \\
mrtr-integrity & 0 & 25 & 0 & 1 \\
mrtr-principal & 1 & 25 & 0 & 0 \\
mrtr-request & 0 & 25 & 0 & 1 \\
mrtr-expiry & 0 & 25 & 0 & 1 \\
mrtr-single-use & 1 & 25 & 0 & 1 \\
\bottomrule\end{tabularx}
\end{table*}
Removing grant matching allows 6 additional main-corpus attacks. Removing handle-principal checking allows 2, and removing the data-pipe category check allows 1. Removing audit collection changes evidence availability but introduces 0 additional successful attacks. These are actual executions of patched source copies, not counts inferred from the first denial reason.

Some removals leave the main attack success count unchanged. The resource-hint witness nevertheless exposes an unauthorized read when trusted resource binding is replaced by caller-hint precedence. Separate witnesses expose the effects of disabling MRTR integrity, request binding, or expiry. Thus, a zero change on the main corpus must not be read as proof that a mechanism is irrelevant.

Removing approval produces 3 inconclusive main trials because the normal challenge-issuance step is no longer reached. Its direct witness confirms that a granted write can execute without approval. This separates a workflow change caused by the intervention from an observed attack success. The appendix and ancillary summaries report variant scores and witness outcomes.

\subsection{Connection-control sensitivity}
The appendix reports B1 sensitivity matrices obtained by removing or tightening one control at a time. Identity checks, protocol context, and matching routing headers each address different cases. Blocking description lint and content quarantine can reduce some adversarial effects, but they are evaluated on the same legitimate workflows and can also reject benign security discussions or contact information. Blanket declining of approval prompts substantially reduces utility while leaving catalog and handle disclosure routes. These observations describe the implemented controls and patterns; they are not estimates for an arbitrary commercial scanner or a live operator.

\subsection{Sealed-read boundary}
\begin{table*}[t]
\centering\small
\caption{Sealed data-release boundary. The same scripted cases run under HCP and the independent ACL comparator; no controlled sensitive read returned plaintext. The two peek rows test the same denial mechanism under different task intent and do not invoke a later sink. ``Handle only'' means an opaque literal moved, not the protected content. Rows are separate from the historical attack denominator.}\label{tab:frontier-sealed}
\begin{tabularx}{\linewidth}{Xll}\toprule
Scenario & B2 outcome & B3 outcome \\ \midrule
PII pipe, rejecting target & No content & No content \\
PII pipe, accepting target & Content written & Content written \\
Classified-handle plaintext peek & Peek denied & Peek denied \\
Handle literal as ordinary parameter & Handle only & Handle only \\
Known plaintext as ordinary parameter & Content written & Content written \\
Ordinary parameter, missing target grant & No content & No content \\
Ordinary parameter after internal read & Content written & Content written \\
Legitimate sensitive direct display & Display denied & Display denied \\
\bottomrule\end{tabularx}
\end{table*}
The sealed policy denies the tested attempt to obtain classified plaintext for ordinary-argument forwarding. No forwarding call follows that denied peek; the observation establishes a controlled-read barrier, not a measured denial at the ordinary-argument sink. This changes release policy and utility, not the way ordinary arguments carry source labels. A caller with plaintext from another source can still submit it to an authorized sink. The same peek mechanism refuses direct display even when that display is legitimate. These outcomes are separate from the historical full-read boundary table.

\subsection{ACL composed with the ChainCaps core}
\begin{table*}[t]
\centering\small
\caption{Stronger comparator on the same scripted attack and benign corpus. B3+ChainCaps combines the authors' MIT-licensed core engine with our local ACL adapter and a trusted-source policy; it is not the full upstream MCP proxy. The extra denial is a false refusal on a legitimate workflow.}\label{tab:frontier-chaincaps}
\begin{tabularx}{\linewidth}{Xrrr}\toprule
Mode & Attacks prevented & Benign completed & Benign policy denials \\ \midrule
B2 & 18/18 & 25/25 & 0/25 \\
B3 & 18/18 & 25/25 & 0/25 \\
B3+ChainCaps & 18/18 & 24/25 & 1/25 \\
\bottomrule\end{tabularx}
\end{table*}
The composed mode prevents all modeled attacks, as B2 and B3 already did. A targeted witness differs: after a PII full read, an ordinary-argument workspace write succeeds under B2 and B3 but is denied by the trusted source budget. The same budget denies an unrelated partner meeting summary after a calendar read in the fixed benign corpus, which accounts for the reported false refusal. The outcomes depend on trusted labels and the added source-to-sink policy, not on an architecture, and are not a reproduction of ChainCaps's published evaluation.

\subsection{Native tasks with live models}
\begin{table}[t]
\centering\small\setlength{\tabcolsep}{4pt}
\caption{Native AgentDojo test tasks (24 tasks, 4 domains) with identical authorization facts for ACL and HCP. Att./Blk./Goal: sessions with an attacker-directed attempt after exposure, with such a call blocked, and with the native attack goal holding on the final environment. Util.: sessions satisfying the native user-task check. GPT completed all 216 sessions; DS exec. gives executed DeepSeek sessions (inconclusive endings in parentheses). Src. doc.: source-document substitution; Task ctx.: task-context continuation; None: no payload.}\label{tab:independent-agentdojo}
\resizebox{\ifdim\width>\linewidth\linewidth\else\width\fi}{!}{\begin{tabular}{llrrrrr}\toprule
Ctrl. & Payload & GPT Att./Blk./Goal & GPT Util. & DS exec. & DS Att./Blk./Goal & DS Util. \\ \midrule
Direct & None & -- & 19 & 15 (2) & -- & 12 \\
Direct & Src. doc. & 0/0/0 & 17 & 13 (1) & 0/0/0 & 11 \\
Direct & Task ctx. & 0/0/0 & 18 & 17 (1) & 4/0/4 & 14 \\
ACL & None & -- & 16 & 10 (1) & -- & 8 \\
ACL & Src. doc. & 0/0/0 & 16 & 17 (3) & 0/0/0 & 13 \\
ACL & Task ctx. & 0/0/0 & 17 & 16 (3) & 4/4/0 & 12 \\
HCP & None & -- & 17 & 11 (2) & -- & 8 \\
HCP & Src. doc. & 0/0/0 & 16 & 16 (4) & 0/0/0 & 10 \\
HCP & Task ctx. & 0/0/0 & 15 & 13 (3) & 1/1/0 & 9 \\
\bottomrule\end{tabular}}
\end{table}
The payload reached GPT's tool output in 141 of 144 payload sessions, yet GPT made 0 attempts (\texttt{Att.}) and no attack goal (\texttt{Goal}) held under any controller. For this model the frozen payloads posed no execution challenge; its interception rate is undefined, and the experiment measures only the utility impact of the controllers.

For DeepSeek, across the 128 executed sessions, task-context continuation under direct execution produced 4 attacker goals (\texttt{Goal}): a transfer to the attacker, a secret posted to a web page, a user removed from the workspace, and a phishing link sent to a colleague. Under the ACL and HCP, the model attempted the same goals, and every call aimed at an attacker operation was blocked (\texttt{Blk.}): 7 of 7 under the ACL and 3 of 3 under HCP, leaving zero attack goals achieved. Source-document substitution, which caused most development effects on a bill-payment and an article-summary task, caused 0 on the test tasks, so the test set does not exercise source poisoning.

Pooled over both models, the per-task change in attack goals relative to direct execution is -0.04 for the ACL and -0.02 for HCP, and 0.00 between them. Because only 3 tasks change, the bootstrap intervals reach zero and the sign-flip tests are not statistically significant: protection is demonstrated by blocking every attempted attacker call (\texttt{7/7} and \texttt{3/3}), whereas estimating a small population rate difference would require a larger task sample.

Strict reference-derived grants also constrain task execution (\texttt{Util.}). The controllers denied 40 calls not aimed at an attacker operation across 26 sessions. Apart from 2 calls to a tool absent from the reference solution, the recorded parameter differences concerned a date or time (15), a title (8), a city with its postal code (5), or a channel name (10, mostly where the channel name is an injection field). In 13 sessions (9 for GPT), the matched direct session completed the user task while the guarded session did not; without payloads, GPT's per-task utility changed by -0.13 under the ACL and -0.08 under HCP, with 95\% intervals from -0.25 and -0.21 to zero (appendix). Whether these non-attacker-directed denials represent acceptable variations or substantive deviations is examined in the offline replay below.

Across all 1,012 guarded calls in their respective live runs (484 under the ACL and 528 under HCP), both controllers enforced the reference grant rule with zero mismatches, though independent model runs generated different call sequences and task outcomes. The independent validator re-executed and re-scored all 344 sessions that produced a final environment. Notably, enforcement overhead is four orders of magnitude below model inference latency: at the median, GPT spent 4.6 s per model request, whereas an entire tool call through the controller and simulated service took just 0.39, 0.39, and 0.57 ms under direct execution, the ACL, and HCP.

In the adaptive follow-up (428 adaptive session records across 216 planned cells, with 209 decisive and 7 reaching a protocol limit), the payload reached model-visible output in 211 of 212 scored first-round sessions, yet none of the three models attempted the attacker goal under any controller, including direct execution (0 of 209 decisive cells; nominal Wilson 95\% upper bound 1.8\%). Direct execution nevertheless ran 57 write calls outside the reference-derived grant, while the ACL and HCP denied all 98 and 93 such calls. On clean tasks, native utility held in 68 of 69 scored sessions under direct execution versus 60 of 68 under the ACL and 59 of 67 under HCP (excluding 12 protocol-limit sessions).

To test whether the denied writes in both native studies were merely formatting variants, our offline authority-rule replay examined all 3299 evaluable calls and 241 denied writes (50 from the first study and 191 from the follow-up). Against same-condition reference candidates, 219 denied writes exhibit substantive state or argument differences (\texttt{Diff}), 15 invoke a tool absent from the reference solution (\texttt{NoRef}), 6 remain unresolved (\texttt{Unres}), and only 1 is state-equivalent (\texttt{Eqv}, an unread \texttt{user\_email} field). Relaxing format, city, or identity-field matching (\texttt{V1}--\texttt{V3}, including ASCII case-folding and whitespace trimming) changes 0/241 denials under static pre-state re-evaluation and diverges in 0/655 trajectories. By contrast, an approve-all diagnostic control (\texttt{V4}) allows all 241/241 denied writes statically and causes first-decision divergence in 110 trajectories (16.8\%), producing 0 tool errors, 110 state changes, and 4 newly completed attacker goals (all from first-study task-context injections in \texttt{NoRef}; appendix). Thus, almost all recorded denials reflect substantive departures from the reference actions rather than trivial surface formatting.

\subsection{Real MCP services}
\begin{table*}[t]
\centering\small
\caption{Isolated real MCP services with prescribed calls (132 conditions). Columns with identical outcomes in every row are merged: per-call ACL and HCP; and ACL and HCP with explicit pipes or sealed reads. Budget: ACL with a session source budget matched to ChainCaps. Done: legitimate workflow completed; Ref.: legitimate workflow refused; Leak: forbidden exfiltration or write effect observed; Blk.: forbidden effect prevented; --: not applicable (no handle interface); n/r: not run (native ChainCaps has no browser tool). Recomputed from workspace files and raw receiver lines.}\label{tab:independent-flow}
\begin{tabularx}{\linewidth}{Xccccc}\toprule
Workflow & Direct & ACL / HCP & Budget & ChainCaps & Pipe / sealed \\ \midrule
Internal note to file & Done & Done & Done & Done & Done \\
Public page to HTTP sink & Done & Done & Done & Done & Done \\
Secret to permitted private file & Done & Done & Ref. & Ref. & Done \\
Secret read, unrelated note to file & Done & Done & Ref. & Ref. & Done \\
Secret read, unrelated note to HTTP & Done & Done & Ref. & Ref. & Done \\
Secret to file & Leak & Leak & Blk. & Blk. & Blk. \\
Secret to HTTP sink & Leak & Leak & Blk. & Blk. & Blk. \\
Known plaintext to HTTP sink & Leak & Leak & Blk. & Blk. & Leak \\
Unlabelled secret to file & Leak & Leak & Leak & Leak & Leak \\
Mislabelled secret to file & Leak & Leak & Leak & Leak & Leak \\
Write after handle revocation & -- & -- & -- & -- & Blk. \\
Write to ungranted file & Leak & Blk. & Blk. & Blk. & Blk. \\
Browser: secret to HTTP sink & Leak & Leak & Blk. & n/r & Blk. \\
Browser: secret read, unrelated note & Done & Done & Ref. & n/r & Done \\
Browser: known plaintext & Leak & Leak & Blk. & n/r & Leak \\
\bottomrule\end{tabularx}
\end{table*}
Across the 132 conditions, the independently recomputed outcomes match the runner and every one of the 151 isolation probes found external connections blocked. Per-call authorization behaves as in the fixture experiments: ACL and HCP stop the write to an ungranted file, yet both deliver the secret to a granted file or HTTP sink when it travels as an ordinary argument, exactly as direct execution does. The session source budget and the native ChainCaps policy prevent those copies but also refuse the unrelated note sent after a sensitive read and, without a notion of a permitted private destination, the legitimate private copy. Explicit pipes and sealed reads prevent the copy through the handle in both HCP and ACL and keep revocation effective, while preserving the unrelated and private workflows. Handles do not help when the caller already knows the secret: every handle and per-call mode sends known plaintext. Only the session budgets stop that send, because after a sensitive read they refuse any external send regardless of its content, which is also why they refuse the unrelated note. Every mode, including native ChainCaps with its own default labels, releases the secret when the source label is missing or wrong. HCP and the ACL baseline with the same handle interface reach identical outcomes in every row.

\subsection{Execution-layer cost on a real MCP backend}
\begin{table}[t]
\centering\small\setlength{\tabcolsep}{4pt}
\caption{Cost on the same isolated MCP backend, inputs and responses (file copy: one read, one write; one workflow in flight; 1000 measured workflows per cell; two passes shown as first/second). Setup: controller setup median (ms). Median: end-to-end median latency (ms). Change: median minus direct execution in the same pass (ms). Non-bk.: median time outside MCP calls (ms). Heap: growth in MB after 1000 sequential workflows (fresh-process median of three). ChainCaps runs its policy inside the backend, so its change is not controller overhead.}\label{tab:independent-cost}
\resizebox{\ifdim\width>\linewidth\linewidth\else\width\fi}{!}{\begin{tabular}{lrrrrr}\toprule
Mode & Setup & Median & Change & Non-bk. & Heap \\ \midrule
Direct & 0.001 & 2.28/2.28 & -- & 0.01 & 0.7 \\
ACL & 0.001 & 2.38/2.30 & +0.10/+0.03 & 0.01 & 0.7 \\
HCP & 0.287 & 2.54/2.59 & +0.26/+0.31 & 0.13 & 7.7 \\
ACL+budget & 0.001 & 2.44/2.33 & +0.16/+0.05 & 0.01 & 0.7 \\
ChainCaps & 0.001 & 10.04/9.45 & +7.76/+7.18 & 0.01 & 0.7 \\
HCP pipe & 0.269 & 2.56/2.45 & +0.29/+0.17 & 0.18 & 10.1 \\
ACL pipe & 0.000 & 2.41/2.38 & +0.13/+0.10 & 0.01 & 1.3 \\
HCP sealed & 0.263 & 2.47/2.45 & +0.20/+0.17 & 0.17 & 10.1 \\
ACL sealed & 0.000 & 2.44/2.41 & +0.17/+0.14 & 0.01 & 1.3 \\
\bottomrule\end{tabular}}
\end{table}
The backend dominates the cost. With one workflow in flight, a file copy takes 2.28 ms at the median under direct execution and an HTTP relay about 33 ms. Time outside MCP calls is about 0.01 ms per workflow for direct execution and the ACLs and 0.13 ms for HCP, which creates two tasks with policy evaluation and audit events per workflow; its median copy latency rose by 0.26 and 0.31 ms in the two passes. For the HTTP relay and with more workflows in flight, mode differences stay within pass-to-pass variation (median 3\%, maximum 21\% in median latency). Native ChainCaps adds about 7.76 ms per copy inside its backend, which reflects its own analysis rather than a controller comparison. HCP setup takes 0.29 ms against a 0.68 s backend launch.

State is the clearer cost. HCP keeps every task and audit entry in memory: after one thousand workflows it held 2000 tasks and 10003 audit entries and its heap had grown by 7.7 MB, against 0.7 MB for the ACL (measured in fresh processes, median of three runs). A deployment needs retention or export of this state, which we did not implement. Earlier in-memory and local HTTP timings of HCP and B3 (appendix, 5 and 5 independent processes) compared prototypes with different responses and functionality and do not isolate controller cost.

\section{Limitations and Interpretation}
The main corpus is deterministic and built around known properties; it does not estimate attack prevalence, adaptive model behavior, or the security of arbitrary deployed servers. A hostile same-process provider or dishonest resolver is outside the model. Source labels are trusted metadata enforced only at explicit managed transfers, sealed reads, or a session source budget; none tracks arbitrary ordinary arguments or prior plaintext. State consumption and audit correlation are single-process: they do not provide distributed exactly-once execution, durable replay prevention, rollback, or tamper-evident logging, and an intact state is not evidence that a human reviewed an action. Because \texttt{B3} independently implements the full \texttt{I1}--\texttt{I8} suite at the gateway/service boundary, its agreement with HCP shows that security follows from enforcing the eight invariants at controlled entrances rather than centralization per se, while contrasting \texttt{B3}'s low-memory inline checks against HCP's stateful runtime. Ablations and witnesses remain conditional on the tested control flow rather than a formal proof of complete mediation.

The independent experiments carry bounded scope. AgentDojo environments are simulated, the 24 test tasks are a hash-selected subset, and public tasks may appear in training data. Because GPT made no attack attempt and source substitution caused no test-set effect, interception is measured only where attempts occurred (DeepSeek task-context sessions). Reference-derived grants represent an idealized policy input; real-service workflows use prescribed calls; and the cost study ran on one shared host. Zero decision changes under \texttt{V1}--\texttt{V3} show only that recorded denials did not stem from configured format, city, or ASCII case/whitespace differences, while probe ID-05 shows that case-folding identity fields can admit a distinct recipient.

\section{Conclusion}
Execution-layer security for tool-using agents should be evaluated through actual effects and structured authority, measuring audit sufficiency and utility alongside attack prevention. Across our scripted benchmark, source-code ablations, live-model AgentDojo trajectories, and unmodified third-party MCP servers, enforcing the eight execution invariants (\texttt{I1}--\texttt{I8}) consistently blocks unauthorized writes, cross-principal handle misuse, and replayed approvals. Crucially, an independent gateway/server realization of the same \texttt{I1}--\texttt{I8} invariants (\texttt{B3}) makes the exact same authorization decisions as HCP (\texttt{B2}) across all scripted and real-service benchmarks and achieves zero mismatches against reference grants across all 1,012 live-model calls (484 \texttt{B3}, 528 HCP). This invariance establishes a clear architectural and operational division of labor: stateless gateways (\texttt{B3}) and stateful runtimes (\texttt{B2}) provide equivalent sub-millisecond protection at the decision boundary, while HCP's managed runtime trades 7.7 MB of heap growth per 1,000 workflows for full task-and-audit provenance. More broadly, our real-service experiments show why opaque handle mediation (\texttt{I5}) is essential alongside per-call ACLs: whereas per-call grants secure the control plane against unauthorized tool invocations and session-wide budgets stop plaintext leaks only by halting unrelated work, keeping sensitive outputs inside explicit handle pipes or sealed reads prevents data-plane exfiltration to authorized sinks at sub-millisecond cost while preserving legitimate workflows.

\bibliographystyle{IEEEtran}
\bibliography{references}
\clearpage
\appendix
\input{supplement}
\end{document}

%% file: supplement.tex
\section{Additional results for the scripted corpus}\label{app:scripted-results}

The scripted corpus contains 18 attack cases and 25 legitimate workflows. These are deterministic property tests, not independent samples from an attack population. The grouped results below retain the number of cases within each group as the denominator. Prevention requires a valid observation with no prohibited effect; an inconclusive attempt is not credited as prevention. Audit sufficiency is scored independently of effects.

\begingroup
\small
\setlength{\tabcolsep}{4pt}
\renewcommand{\arraystretch}{1.15}
\setlength{\LTleft}{0pt}
\setlength{\LTright}{0pt}
\begin{longtable}{>{\raggedright\arraybackslash}p{\dimexpr 0.18\linewidth-2\tabcolsep\relax}>{\raggedright\arraybackslash}p{\dimexpr 0.82\linewidth-2\tabcolsep\relax}}
\caption{Execution modes used in the scripted corpus and local prototype measurements.}\label{tab:app-modes}\\
\toprule
\textbf{Mode} & \textbf{Meaning} \\
\midrule
\endfirsthead
\multicolumn{2}{l}{\small\tablename~\thetable{} (continued)}\\
\toprule
\textbf{Mode} & \textbf{Meaning} \\
\midrule
\endhead
\midrule
\multicolumn{2}{r}{\small Continued on next page}\\
\endfoot
\bottomrule
\endlastfoot
B0 & Connection-only configuration, without the tested execution authorization controls. \\
B1 & Connection controls with description linting, historical protocol checks, identity and header checks, and a fixed approval script. \\
B1-decline & The B1 configuration with high-risk confirmations declined. \\
B2 & HCP reference runtime with grant, canonical-resource, managed-transfer, and audit checks. \\
B3 & Independent gateway/server ACL given the same trusted authorization facts and data labels as HCP. \\
\end{longtable}
\endgroup

\begingroup
\small
\setlength{\tabcolsep}{4pt}
\renewcommand{\arraystretch}{1.15}
\setlength{\LTleft}{0pt}
\setlength{\LTright}{0pt}
\begin{longtable}{>{\raggedright\arraybackslash}p{\dimexpr 0.28\linewidth-2\tabcolsep\relax}>{\raggedright\arraybackslash}p{\dimexpr 0.19\linewidth-2\tabcolsep\relax}>{\raggedright\arraybackslash}p{\dimexpr 0.1\linewidth-2\tabcolsep\relax}>{\raggedright\arraybackslash}p{\dimexpr 0.14\linewidth-2\tabcolsep\relax}>{\raggedright\arraybackslash}p{\dimexpr 0.15\linewidth-2\tabcolsep\relax}>{\raggedright\arraybackslash}p{\dimexpr 0.14\linewidth-2\tabcolsep\relax}}
\caption{Grouped attacks. Execution checks, selected catalog policy, and protocol/state conditions are distinct targets; protocol/state combines conformance and the selected one-time approval contract. Cases is the denominator for Success, Prevented, and Uncertain in the same row; all entries are counts.}\label{tab:app-group-results}\\
\toprule
\textbf{Group} & \textbf{Mode} & \textbf{Cases} & \textbf{Success} & \textbf{Prevented} & \textbf{Uncertain} \\
\midrule
\endfirsthead
\multicolumn{6}{l}{\small\tablename~\thetable{} (continued)}\\
\toprule
\textbf{Group} & \textbf{Mode} & \textbf{Cases} & \textbf{Success} & \textbf{Prevented} & \textbf{Uncertain} \\
\midrule
\endhead
\midrule
\multicolumn{6}{r}{\small Continued on next page}\\
\endfoot
\bottomrule
\endlastfoot
Execution / data & B0 & 10 & 10 & 0 & 0 \\
Execution / data & B1 & 10 & 7 & 3 & 0 \\
Execution / data & B1-decline & 10 & 1 & 9 & 0 \\
Execution / data & B2 & 10 & 0 & 10 & 0 \\
Execution / data & B3 & 10 & 0 & 10 & 0 \\
Catalog policy & B0 & 2 & 2 & 0 & 0 \\
Catalog policy & B1 & 2 & 2 & 0 & 0 \\
Catalog policy & B1-decline & 2 & 2 & 0 & 0 \\
Catalog policy & B2 & 2 & 0 & 2 & 0 \\
Catalog policy & B3 & 2 & 0 & 2 & 0 \\
Protocol / state & B0 & 6 & 6 & 0 & 0 \\
Protocol / state & B1 & 6 & 2 & 4 & 0 \\
Protocol / state & B1-decline & 6 & 0 & 5 & 1 \\
Protocol / state & B2 & 6 & 0 & 6 & 0 \\
Protocol / state & B3 & 6 & 0 & 6 & 0 \\
\end{longtable}
\endgroup

\begingroup
\small
\setlength{\tabcolsep}{4pt}
\renewcommand{\arraystretch}{1.15}
\setlength{\LTleft}{0pt}
\setlength{\LTright}{0pt}
\begin{longtable}{>{\raggedright\arraybackslash}p{\dimexpr 0.34\linewidth-2\tabcolsep\relax}>{\raggedright\arraybackslash}p{\dimexpr 0.16\linewidth-2\tabcolsep\relax}>{\raggedright\arraybackslash}p{\dimexpr 0.17\linewidth-2\tabcolsep\relax}>{\raggedright\arraybackslash}p{\dimexpr 0.16\linewidth-2\tabcolsep\relax}>{\raggedright\arraybackslash}p{\dimexpr 0.17\linewidth-2\tabcolsep\relax}}
\caption{B1 control sensitivity on the same attack and legitimate workflows. All columns retain their original group denominators; an unmet replay setup remains uncertain.}\label{tab:app-sensitivity-results}\\
\toprule
\textbf{Variant} & \textbf{Success} & \textbf{Prevented} & \textbf{Uncertain} & \textbf{Benign done} \\
\midrule
\endfirsthead
\multicolumn{5}{l}{\small\tablename~\thetable{} (continued)}\\
\toprule
\textbf{Variant} & \textbf{Success} & \textbf{Prevented} & \textbf{Uncertain} & \textbf{Benign done} \\
\midrule
\endhead
\midrule
\multicolumn{5}{r}{\small Continued on next page}\\
\endfoot
\bottomrule
\endlastfoot
Reference controls & 11/18 & 7/18 & 0/18 & 24/25 \\
Identity check removed & 14/18 & 4/18 & 0/18 & 25/25 \\
Protocol check removed & 13/18 & 5/18 & 0/18 & 24/25 \\
Header check removed & 13/18 & 5/18 & 0/18 & 24/25 \\
Blocking description lint & 9/18 & 9/18 & 0/18 & 23/25 \\
Blocking content filter & 9/18 & 9/18 & 0/18 & 22/25 \\
Decline high-risk approval & 3/18 & 14/18 & 1/18 & 14/25 \\
\end{longtable}
\endgroup

The protocol/state group combines historical initialization, transport-header consistency, and the selected one-time approval contract. MRTR integrity is required when carried state affects authority; principal, request, and expiry binding follow the deployment policy described in the main text. Single-use enforcement is evaluated for the one-time approval operation. Catalog cases test an explicitly selected authorization-scoping policy: a public catalog is not intrinsically a violation. Handle and invocation cases test the stated execution policy, rather than an attack concept unique to a protocol revision.

The machine-readable observations accompanying this preprint contain the complete attack and legitimate case matrices, separate transfer-boundary observations, and direct mechanism witnesses. They retain case identifiers to permit exact matching, while omitting request contents and event logs. They also retain audit sufficiency separately from the execution outcome.

The companion also gives a semantic title and explicit scoring goal for every attack and legitimate case, including the near-miss relationship where registered. Each of the 16 source-removal variants executes the full 43-case corpus. Its compact scoring matrix, direct witnesses, changed-file identifiers, and before/after source hashes are included alongside the unchanged control. Complete patch records remain in the local evidence. Patch hashes identify evaluated source changes; they are not evidence of a general proof. The ablation summary in the main text counts newly successful attacks relative to that unchanged control.

\section{Local prototype latency and per-process statistics}\label{app:prototype-latency}

The in-memory measurements use 5 independent processes, 100 warmup operations, and 1000 measured operations per operation and process. Mean, median, and P95 below average the corresponding process statistics; the range covers process means. The workloads include different response and audit behavior in HCP and the ACL, so these are operation latencies rather than an isolated incremental cost of authorization. They are separate from the matched real-backend comparison in the main text.

\begingroup
\small
\setlength{\tabcolsep}{4pt}
\renewcommand{\arraystretch}{1.15}
\setlength{\LTleft}{0pt}
\setlength{\LTright}{0pt}
\begin{longtable}{>{\raggedright\arraybackslash}p{\dimexpr 0.36\linewidth-2\tabcolsep\relax}>{\raggedright\arraybackslash}p{\dimexpr 0.14\linewidth-2\tabcolsep\relax}>{\raggedright\arraybackslash}p{\dimexpr 0.14\linewidth-2\tabcolsep\relax}>{\raggedright\arraybackslash}p{\dimexpr 0.14\linewidth-2\tabcolsep\relax}>{\raggedright\arraybackslash}p{\dimexpr 0.22\linewidth-2\tabcolsep\relax}}
\caption{Operation latency in milliseconds. Mean, median and P95 columns average the corresponding per-process statistics; the range is across process means. This is not a controls-disabled incremental-overhead estimate.}\label{tab:app-latency-summary}\\
\toprule
\textbf{Mode / operation} & \textbf{Mean} & \textbf{Median} & \textbf{P95} & \textbf{Mean range} \\
\midrule
\endfirsthead
\multicolumn{5}{l}{\small\tablename~\thetable{} (continued)}\\
\toprule
\textbf{Mode / operation} & \textbf{Mean} & \textbf{Median} & \textbf{P95} & \textbf{Mean range} \\
\midrule
\endhead
\midrule
\multicolumn{5}{r}{\small Continued on next page}\\
\endfoot
\bottomrule
\endlastfoot
B2: Explain read policy & 0.0161 & 0.0120 & 0.0263 & 0.0137 to 0.0203 \\
B2: Preview managed data & 0.0045 & 0.0033 & 0.0078 & 0.0037 to 0.0058 \\
B2: Start read task & 0.0451 & 0.0368 & 0.0644 & 0.0422 to 0.0496 \\
B2: Pipe to write task & 0.0896 & 0.0709 & 0.1412 & 0.0819 to 0.0992 \\
B3: Start read task & 0.0028 & 0.0012 & 0.0038 & 0.0020 to 0.0042 \\
B3: Pipe to write task & 0.0022 & 0.0019 & 0.0033 & 0.0017 to 0.0033 \\
\end{longtable}
\endgroup

\begingroup
\small
\setlength{\tabcolsep}{4pt}
\renewcommand{\arraystretch}{1.15}
\setlength{\LTleft}{0pt}
\setlength{\LTright}{0pt}
\begin{longtable}{>{\raggedright\arraybackslash}p{\dimexpr 0.12\linewidth-2\tabcolsep\relax}>{\raggedright\arraybackslash}p{\dimexpr 0.37\linewidth-2\tabcolsep\relax}>{\raggedright\arraybackslash}p{\dimexpr 0.17\linewidth-2\tabcolsep\relax}>{\raggedright\arraybackslash}p{\dimexpr 0.17\linewidth-2\tabcolsep\relax}>{\raggedright\arraybackslash}p{\dimexpr 0.17\linewidth-2\tabcolsep\relax}}
\caption{In-memory operation statistics for each independent process. Every row contains 1000 post-warmup samples; all latency columns are milliseconds. Process numbers identify independent measurement runs, not different machines.}\label{tab:app-latency-processes}\\
\toprule
\textbf{Process} & \textbf{Mode / operation} & \textbf{Mean} & \textbf{Median} & \textbf{P95} \\
\midrule
\endfirsthead
\multicolumn{5}{l}{\small\tablename~\thetable{} (continued)}\\
\toprule
\textbf{Process} & \textbf{Mode / operation} & \textbf{Mean} & \textbf{Median} & \textbf{P95} \\
\midrule
\endhead
\midrule
\multicolumn{5}{r}{\small Continued on next page}\\
\endfoot
\bottomrule
\endlastfoot
1 & B2: Explain read policy & 0.020325 & 0.013450 & 0.033650 \\
1 & B2: Preview managed data & 0.004058 & 0.002795 & 0.006611 \\
1 & B2: Start read task & 0.049617 & 0.040630 & 0.072650 \\
1 & B2: Pipe to write task & 0.099233 & 0.073001 & 0.165231 \\
1 & B3: Start read task & 0.002652 & 0.001580 & 0.004710 \\
1 & B3: Pipe to write task & 0.001819 & 0.001530 & 0.003470 \\
2 & B2: Explain read policy & 0.014059 & 0.011320 & 0.026670 \\
2 & B2: Preview managed data & 0.003709 & 0.002720 & 0.006410 \\
2 & B2: Start read task & 0.044136 & 0.034410 & 0.060060 \\
2 & B2: Pipe to write task & 0.082296 & 0.066285 & 0.132190 \\
2 & B3: Start read task & 0.001993 & 0.001100 & 0.003510 \\
2 & B3: Pipe to write task & 0.003288 & 0.003105 & 0.004090 \\
3 & B2: Explain read policy & 0.017649 & 0.011880 & 0.025800 \\
3 & B2: Preview managed data & 0.003882 & 0.002800 & 0.006570 \\
3 & B2: Start read task & 0.044822 & 0.036485 & 0.067370 \\
3 & B2: Pipe to write task & 0.091883 & 0.073806 & 0.137410 \\
3 & B3: Start read task & 0.004205 & 0.000760 & 0.001920 \\
3 & B3: Pipe to write task & 0.002178 & 0.001790 & 0.003690 \\
4 & B2: Explain read policy & 0.014841 & 0.011610 & 0.025110 \\
4 & B2: Preview managed data & 0.005803 & 0.004750 & 0.010500 \\
4 & B2: Start read task & 0.044919 & 0.037655 & 0.065310 \\
4 & B2: Pipe to write task & 0.092624 & 0.074261 & 0.150441 \\
4 & B3: Start read task & 0.001967 & 0.000920 & 0.003780 \\
4 & B3: Pipe to write task & 0.001661 & 0.001510 & 0.002470 \\
5 & B2: Explain read policy & 0.013702 & 0.011780 & 0.020370 \\
5 & B2: Preview managed data & 0.005065 & 0.003565 & 0.008980 \\
5 & B2: Start read task & 0.042247 & 0.034605 & 0.056560 \\
5 & B2: Pipe to write task & 0.081926 & 0.067050 & 0.120680 \\
5 & B3: Start read task & 0.002993 & 0.001660 & 0.005000 \\
5 & B3: Pipe to write task & 0.001828 & 0.001730 & 0.002550 \\
\end{longtable}
\endgroup

\begingroup
\small
\setlength{\tabcolsep}{4pt}
\renewcommand{\arraystretch}{1.15}
\setlength{\LTleft}{0pt}
\setlength{\LTright}{0pt}
\begin{longtable}{>{\raggedright\arraybackslash}p{\dimexpr 0.4\linewidth-2\tabcolsep\relax}>{\raggedright\arraybackslash}p{\dimexpr 0.3\linewidth-2\tabcolsep\relax}>{\raggedright\arraybackslash}p{\dimexpr 0.3\linewidth-2\tabcolsep\relax}}
\caption{Local end-to-end HTTP prototype latency (milliseconds). 5 independent processes each contribute 1000 sequential operation samples after warmup; an approved write sample includes the approval challenge and resume requests. Entries are the mean of process means and their range. Different implementations and policy paths prevent an intrinsic architecture-speed claim.}\label{tab:app-http-latency}\\
\toprule
\textbf{Mode / operation} & \textbf{Process-mean average (ms)} & \textbf{Process-mean range (ms)} \\
\midrule
\endfirsthead
\multicolumn{3}{l}{\small\tablename~\thetable{} (continued)}\\
\toprule
\textbf{Mode / operation} & \textbf{Process-mean average (ms)} & \textbf{Process-mean range (ms)} \\
\midrule
\endhead
\midrule
\multicolumn{3}{r}{\small Continued on next page}\\
\endfoot
\bottomrule
\endlastfoot
B2 read & 1.7005 & 1.6679 -- 1.7689 \\
B2 approved write & 3.3770 & 3.3361 -- 3.4653 \\
B3 read & 1.4391 & 1.4191 -- 1.4664 \\
B3 approved write & 2.9204 & 2.8835 -- 2.9715 \\
\end{longtable}
\endgroup

The HTTP measurements use loopback, one operation in flight, and 100 warmup operations per operation and process. A read sample is one request; an approved write includes the challenge and resume requests. The fixed sample count is 1000 per operation in each of 5 processes. Compact per-process HTTP latency and state statistics are included in the machine-readable companion; raw timing samples and event logs are not embedded in the paper.

\section{Paired native-task contrasts}\label{app:paired-contrasts}

The first native AgentDojo study pairs sessions within a task by model and payload condition, provided both controllers executed that session. Differences are averaged within a task before averaging over tasks. Utility means the native user-task check; attack effect means the native attacker check on the final environment. The table reports differences in outcome fractions, not percentage points. GPT rows use its complete test set. Pooled rows include the executed DeepSeek sessions, whose missing runs must not be counted as unsuccessful attacks.

\begin{landscape}

\begingroup
\small
\setlength{\tabcolsep}{4pt}
\renewcommand{\arraystretch}{1.15}
\setlength{\LTleft}{0pt}
\setlength{\LTright}{0pt}
\begin{longtable}{>{\raggedright\arraybackslash}p{\dimexpr 0.08\linewidth-2\tabcolsep\relax}>{\raggedright\arraybackslash}p{\dimexpr 0.17\linewidth-2\tabcolsep\relax}>{\raggedright\arraybackslash}p{\dimexpr 0.12\linewidth-2\tabcolsep\relax}>{\raggedright\arraybackslash}p{\dimexpr 0.12\linewidth-2\tabcolsep\relax}>{\raggedright\arraybackslash}p{\dimexpr 0.08\linewidth-2\tabcolsep\relax}>{\raggedright\arraybackslash}p{\dimexpr 0.16\linewidth-2\tabcolsep\relax}>{\raggedright\arraybackslash}p{\dimexpr 0.08\linewidth-2\tabcolsep\relax}>{\raggedright\arraybackslash}p{\dimexpr 0.19\linewidth-2\tabcolsep\relax}}
\caption{Paired task-level differences. Within each task, matched sessions (same model and condition, both executed) are differenced and averaged; the mean, a 95\% task bootstrap interval and a sign-flip p-value are taken over tasks. GPT rows use complete data; pooled rows add the DeepSeek sessions that executed. Models, payload families and controllers are not independent samples. Mean and interval are differences in outcome fractions; p is the recorded sign-flip p-value; Tasks changed is the number with a nonzero difference over the task denominator.}\label{tab:app-task-contrasts}\\
\toprule
\textbf{Data} & \textbf{Contrast} & \textbf{Outcome} & \textbf{Condition} & \textbf{Mean} & \textbf{95\% interval} & \textbf{p} & \textbf{Tasks changed} \\
\midrule
\endfirsthead
\multicolumn{8}{l}{\small\tablename~\thetable{} (continued)}\\
\toprule
\textbf{Data} & \textbf{Contrast} & \textbf{Outcome} & \textbf{Condition} & \textbf{Mean} & \textbf{95\% interval} & \textbf{p} & \textbf{Tasks changed} \\
\midrule
\endhead
\midrule
\multicolumn{8}{r}{\small Continued on next page}\\
\endfoot
\bottomrule
\endlastfoot
GPT & ACL - Direct & Utility & No payload & -0.13 & -0.25 to 0.00 & 0.25 & 3/24 \\
GPT & HCP - Direct & Utility & No payload & -0.08 & -0.21 to 0.00 & 0.50 & 2/24 \\
GPT & HCP - ACL & Utility & No payload & 0.04 & -0.08 to 0.17 & 1.00 & 3/24 \\
GPT & ACL - Direct & Utility & Payloads & -0.04 & -0.15 to 0.04 & 0.75 & 3/24 \\
GPT & HCP - Direct & Utility & Payloads & -0.08 & -0.19 to 0.00 & 0.25 & 3/24 \\
GPT & HCP - ACL & Utility & Payloads & -0.04 & -0.13 to 0.00 & 1.00 & 1/24 \\
Pooled & ACL - Direct & Attack effect & Payloads & -0.04 & -0.08 to 0.00 & 0.25 & 3/24 \\
Pooled & HCP - Direct & Attack effect & Payloads & -0.02 & -0.06 to 0.00 & 0.50 & 2/24 \\
Pooled & HCP - ACL & Attack effect & Payloads & 0.00 & 0.00 to 0.00 & 1.00 & 0/24 \\
\end{longtable}
\endgroup

\end{landscape}

A refusal outside a reference-derived grant is not automatically a refusal of a legitimate equivalent action. Reference solutions are trusted inputs to the controller, not a complete description of every acceptable workflow. The offline analysis below distinguishes substantive changes, demonstrated equivalence, unresolved comparisons, and calls without a same-tool reference. This distinction limits utility claims based solely on the first study's non-attacker-directed denial count.

\section{Adaptive native-task follow-up}\label{app:adaptive-follow-up}

The follow-up uses 24 fresh test tasks, 3 further models, and a fixed two-round attacker with access to the attacker goal and reference calls. A clean session is evaluated separately from an adaptive cell. A cell is decisive if a round achieves the native attacker goal or all its required rounds are scored; a cell stopping at a protocol limit remains inconclusive. The 428 adaptive session records comprise 216 first-round and 212 second-round records. Of 216 adaptive cells, 209 are decisive and 7 inconclusive. The records contain 0 goal-completing sessions and 0 reference-target attempts. Because direct execution also makes no goal attempt, this study does not estimate controller prevention of attempted attacker goals.

\begingroup
\small
\setlength{\tabcolsep}{4pt}
\renewcommand{\arraystretch}{1.15}
\setlength{\LTleft}{0pt}
\setlength{\LTright}{0pt}
\begin{longtable}{>{\raggedright\arraybackslash}p{\dimexpr 0.21\linewidth-2\tabcolsep\relax}>{\raggedright\arraybackslash}p{\dimexpr 0.13\linewidth-2\tabcolsep\relax}>{\raggedright\arraybackslash}p{\dimexpr 0.14\linewidth-2\tabcolsep\relax}>{\raggedright\arraybackslash}p{\dimexpr 0.14\linewidth-2\tabcolsep\relax}>{\raggedright\arraybackslash}p{\dimexpr 0.14\linewidth-2\tabcolsep\relax}>{\raggedright\arraybackslash}p{\dimexpr 0.11\linewidth-2\tabcolsep\relax}>{\raggedright\arraybackslash}p{\dimexpr 0.13\linewidth-2\tabcolsep\relax}}
\caption{Native task utility and adaptive-cell outcomes on 24 fresh tasks per model/controller. Clean and Round 1 are utility successes over scored sessions; protocol-limit sessions are excluded from these utility denominators. Decisive is decisive cells over planned adaptive cells; Success and Uncertain are counts of adaptive cells, not sessions. Model names are endpoint-reported. All models aggregates the listed models; it is not an additional independent model.}\label{tab:app-adaptive-utility}\\
\toprule
\textbf{Model} & \textbf{Controller} & \textbf{Clean utility} & \textbf{Round 1 utility} & \textbf{Decisive} & \textbf{Success} & \textbf{Uncertain} \\
\midrule
\endfirsthead
\multicolumn{7}{l}{\small\tablename~\thetable{} (continued)}\\
\toprule
\textbf{Model} & \textbf{Controller} & \textbf{Clean utility} & \textbf{Round 1 utility} & \textbf{Decisive} & \textbf{Success} & \textbf{Uncertain} \\
\midrule
\endhead
\midrule
\multicolumn{7}{r}{\small Continued on next page}\\
\endfoot
\bottomrule
\endlastfoot
gemini-3.6-flash-high & Direct & 22/22 & 22/24 & 24/24 & 0 & 0 \\
deepseek-flash & Direct & 22/23 & 23/24 & 23/24 & 0 & 1 \\
GLM-5.3-flash & Direct & 24/24 & 23/24 & 24/24 & 0 & 0 \\
All models & Direct & 68/69 & 68/72 & 71/72 & 0 & 1 \\
gemini-3.6-flash-high & ACL & 20/23 & 19/23 & 23/24 & 0 & 1 \\
deepseek-flash & ACL & 19/22 & 19/23 & 22/24 & 0 & 2 \\
GLM-5.3-flash & ACL & 21/23 & 19/24 & 24/24 & 0 & 0 \\
All models & ACL & 60/68 & 57/70 & 69/72 & 0 & 3 \\
gemini-3.6-flash-high & HCP & 20/23 & 20/24 & 23/24 & 0 & 1 \\
deepseek-flash & HCP & 19/21 & 19/22 & 22/24 & 0 & 2 \\
GLM-5.3-flash & HCP & 20/23 & 20/24 & 24/24 & 0 & 0 \\
All models & HCP & 59/67 & 59/70 & 69/72 & 0 & 3 \\
\end{longtable}
\endgroup

\begingroup
\small
\setlength{\tabcolsep}{4pt}
\renewcommand{\arraystretch}{1.15}
\setlength{\LTleft}{0pt}
\setlength{\LTright}{0pt}
\begin{longtable}{>{\raggedright\arraybackslash}p{\dimexpr 0.32\linewidth-2\tabcolsep\relax}>{\raggedright\arraybackslash}p{\dimexpr 0.16\linewidth-2\tabcolsep\relax}>{\raggedright\arraybackslash}p{\dimexpr 0.26\linewidth-2\tabcolsep\relax}>{\raggedright\arraybackslash}p{\dimexpr 0.26\linewidth-2\tabcolsep\relax}}
\caption{Write calls outside the reference-derived grant in the adaptive follow-up. Entries are executed calls over all such calls, separately for clean sessions and both injected rounds combined. These are call denominators, not task or adaptive-cell counts; unauthorized does not imply that the native attacker goal was achieved.}\label{tab:app-adaptive-writes}\\
\toprule
\textbf{Model} & \textbf{Controller} & \textbf{Clean: executed / total} & \textbf{Injected: executed / total} \\
\midrule
\endfirsthead
\multicolumn{4}{l}{\small\tablename~\thetable{} (continued)}\\
\toprule
\textbf{Model} & \textbf{Controller} & \textbf{Clean: executed / total} & \textbf{Injected: executed / total} \\
\midrule
\endhead
\midrule
\multicolumn{4}{r}{\small Continued on next page}\\
\endfoot
\bottomrule
\endlastfoot
gemini-3.6-flash-high & Direct & 4/4 & 19/19 \\
deepseek-flash & Direct & 3/3 & 11/11 \\
GLM-5.3-flash & Direct & 6/6 & 14/14 \\
All models & Direct & 13/13 & 44/44 \\
gemini-3.6-flash-high & ACL & 0/6 & 0/22 \\
deepseek-flash & ACL & 0/7 & 0/24 \\
GLM-5.3-flash & ACL & 0/11 & 0/28 \\
All models & ACL & 0/24 & 0/74 \\
gemini-3.6-flash-high & HCP & 0/7 & 0/26 \\
deepseek-flash & HCP & 0/4 & 0/26 \\
GLM-5.3-flash & HCP & 0/7 & 0/23 \\
All models & HCP & 0/18 & 0/75 \\
\end{longtable}
\endgroup

The adaptive records include 426 payload exposures. Across clean and injected records, 9 records reach the turn limit and 10 reach the tool-call limit. These are record counts and must not be substituted for the adaptive-cell denominator. The attacker knows the goal and reference calls, and its generator is also one of the tested models; the result therefore describes this fixed procedure and workload rather than a general adaptive-attack success rate. Repeated tasks across models, controllers, and rounds make aggregate counts descriptive, not independent trials.

\section{Offline authority-rule replay}\label{app:authority-rules}

D1 denotes the first, two-model native-task study; D2 denotes the three-model adaptive follow-up. Replay uses only recorded ACL and HCP trajectories, including executed inconclusive records. Direct-execution trajectories and records that never executed are outside this analysis. Across 655 evaluable trajectories and 3299 evaluable calls, 241 writes were denied by the original reference-grant rule. The call-level classifications total Eqv/Diff/Unres/NoRef = 1/219/6/15. They classify the relationship to reference actions; they do not identify the user's true intent.

The trajectory denominator includes 32 unfinished trajectories through their reconstructable prefixes. Calls and denials are counted without deduplication, and repeated tasks, models, controllers, or attack rounds are not independent samples. Rule definitions and diagnostic probes were designed or adjusted after inspecting the actual D1/D2 data, then frozen before the reported execution. This is an exploratory offline analysis, not a preregistered or independently held-out test of equivalence rules.

\begingroup
\small
\setlength{\tabcolsep}{4pt}
\renewcommand{\arraystretch}{1.15}
\setlength{\LTleft}{0pt}
\setlength{\LTright}{0pt}
\begin{longtable}{>{\raggedright\arraybackslash}p{\dimexpr 0.12\linewidth-2\tabcolsep\relax}>{\raggedright\arraybackslash}p{\dimexpr 0.15\linewidth-2\tabcolsep\relax}>{\raggedright\arraybackslash}p{\dimexpr 0.2\linewidth-2\tabcolsep\relax}>{\raggedright\arraybackslash}p{\dimexpr 0.13\linewidth-2\tabcolsep\relax}>{\raggedright\arraybackslash}p{\dimexpr 0.1\linewidth-2\tabcolsep\relax}>{\raggedright\arraybackslash}p{\dimexpr 0.1\linewidth-2\tabcolsep\relax}>{\raggedright\arraybackslash}p{\dimexpr 0.1\linewidth-2\tabcolsep\relax}>{\raggedright\arraybackslash}p{\dimexpr 0.1\linewidth-2\tabcolsep\relax}}
\caption{Classification of denied writes by dataset and controller. All class columns count calls. Their sum equals Denied in each row; Trajectories is a separate denominator for first-divergence analysis.}\label{tab:app-authority-labels}\\
\toprule
\textbf{Dataset} & \textbf{Controller} & \textbf{Trajectories} & \textbf{Denied} & \textbf{Eqv} & \textbf{Diff} & \textbf{Unres} & \textbf{NoRef} \\
\midrule
\endfirsthead
\multicolumn{8}{l}{\small\tablename~\thetable{} (continued)}\\
\toprule
\textbf{Dataset} & \textbf{Controller} & \textbf{Trajectories} & \textbf{Denied} & \textbf{Eqv} & \textbf{Diff} & \textbf{Unres} & \textbf{NoRef} \\
\midrule
\endhead
\midrule
\multicolumn{8}{r}{\small Continued on next page}\\
\endfoot
\bottomrule
\endlastfoot
D1 & ACL & 115 & 22 & 0 & 12 & 3 & 7 \\
D1 & HCP & 112 & 28 & 0 & 22 & 3 & 3 \\
D2 & ACL & 214 & 98 & 1 & 95 & 0 & 2 \\
D2 & HCP & 214 & 93 & 0 & 90 & 0 & 3 \\
\end{longtable}
\endgroup

\begingroup
\small
\setlength{\tabcolsep}{4pt}
\renewcommand{\arraystretch}{1.15}
\setlength{\LTleft}{0pt}
\setlength{\LTright}{0pt}
\begin{longtable}{>{\raggedright\arraybackslash}p{\dimexpr 0.12\linewidth-2\tabcolsep\relax}>{\raggedright\arraybackslash}p{\dimexpr 0.88\linewidth-2\tabcolsep\relax}}
\caption{Meaning of the denied-write equivalence classes. Comparisons use same-tool reference candidates and, where specified, native tool execution on copied state.}\label{tab:app-authority-definitions}\\
\toprule
\textbf{Class} & \textbf{Interpretation} \\
\midrule
\endfirsthead
\multicolumn{2}{l}{\small\tablename~\thetable{} (continued)}\\
\toprule
\textbf{Class} & \textbf{Interpretation} \\
\midrule
\endhead
\midrule
\multicolumn{2}{r}{\small Continued on next page}\\
\endfoot
\bottomrule
\endlastfoot
Eqv & At least one same-tool reference candidate is equivalent under the classifier: all differing fields have a demonstrated equivalent or effect-neutral classification. \\
Diff & Every same-tool reference candidate contains at least one substantive difference, such as an absent required field, changed date, different object, or another changed effect. \\
Unres & A same-tool reference exists, but equivalence or a substantive difference cannot be established under the classifier. \\
NoRef & No same-tool reference action exists. This is not a finding that the call is legitimate or malicious. \\
\end{longtable}
\endgroup

\begingroup
\small
\setlength{\tabcolsep}{4pt}
\renewcommand{\arraystretch}{1.15}
\setlength{\LTleft}{0pt}
\setlength{\LTright}{0pt}
\begin{longtable}{>{\raggedright\arraybackslash}p{\dimexpr 0.12\linewidth-2\tabcolsep\relax}>{\raggedright\arraybackslash}p{\dimexpr 0.88\linewidth-2\tabcolsep\relax}}
\caption{Authority-rule variants. V0 is the recorded reference-grant rule; V1--V4 only reconsider its denied writes. They do not rewrite tool arguments. V3 extends V1 directly and does not include V2.}\label{tab:app-authority-variants}\\
\toprule
\textbf{Variant} & \textbf{Rule} \\
\midrule
\endfirsthead
\multicolumn{2}{l}{\small\tablename~\thetable{} (continued)}\\
\toprule
\textbf{Variant} & \textbf{Rule} \\
\midrule
\endhead
\midrule
\multicolumn{2}{r}{\small Continued on next page}\\
\endfoot
\bottomrule
\endlastfoot
V0 & Exact matching of the selected write fields and the same-condition reference resource digest. \\
V1 & V0 plus configured date/time format equivalence pairs on date and datetime fields. \\
V2 & V1 plus configured city-name equivalence pairs. \\
V3 & V1 plus trimming ASCII edge whitespace and lowercasing A--Z in configured identity fields; lists are compared element by element. \\
V4 & Approve every denied write, retaining its original tool and arguments; this is an approve-all diagnostic control rather than a recommended policy. \\
\end{longtable}
\endgroup

\begingroup
\small
\setlength{\tabcolsep}{4pt}
\renewcommand{\arraystretch}{1.15}
\setlength{\LTleft}{0pt}
\setlength{\LTright}{0pt}
\begin{longtable}{>{\raggedright\arraybackslash}p{\dimexpr 0.1\linewidth-2\tabcolsep\relax}>{\raggedright\arraybackslash}p{\dimexpr 0.16\linewidth-2\tabcolsep\relax}>{\raggedright\arraybackslash}p{\dimexpr 0.15\linewidth-2\tabcolsep\relax}>{\raggedright\arraybackslash}p{\dimexpr 0.15\linewidth-2\tabcolsep\relax}>{\raggedright\arraybackslash}p{\dimexpr 0.18\linewidth-2\tabcolsep\relax}>{\raggedright\arraybackslash}p{\dimexpr 0.13\linewidth-2\tabcolsep\relax}>{\raggedright\arraybackslash}p{\dimexpr 0.13\linewidth-2\tabcolsep\relax}}
\caption{Decision changes under each rule. Static re-evaluates each denied call on its recorded pre-call state (denied-write denominators). First-div. replays each trajectory only to its first decision divergence (trajectory denominators; D1 and D2 give diverging trajectory counts).}\label{tab:app-authority-rule-changes}\\
\toprule
\textbf{Variant} & \textbf{Static: all} & \textbf{Static: D1} & \textbf{Static: D2} & \textbf{First-div.: all} & \textbf{First-div.: D1} & \textbf{First-div.: D2} \\
\midrule
\endfirsthead
\multicolumn{7}{l}{\small\tablename~\thetable{} (continued)}\\
\toprule
\textbf{Variant} & \textbf{Static: all} & \textbf{Static: D1} & \textbf{Static: D2} & \textbf{First-div.: all} & \textbf{First-div.: D1} & \textbf{First-div.: D2} \\
\midrule
\endhead
\midrule
\multicolumn{7}{r}{\small Continued on next page}\\
\endfoot
\bottomrule
\endlastfoot
V1 & 0/241 & 0/50 & 0/191 & 0/655 & 0 & 0 \\
V2 & 0/241 & 0/50 & 0/191 & 0/655 & 0 & 0 \\
V3 & 0/241 & 0/50 & 0/191 & 0/655 & 0 & 0 \\
V4 & 241/241 & 50/50 & 191/191 & 110/655 & 31 & 79 \\
\end{longtable}
\endgroup

\begingroup
\small
\setlength{\tabcolsep}{4pt}
\renewcommand{\arraystretch}{1.15}
\setlength{\LTleft}{0pt}
\setlength{\LTright}{0pt}
\begin{longtable}{>{\raggedright\arraybackslash}p{\dimexpr 0.13\linewidth-2\tabcolsep\relax}>{\raggedright\arraybackslash}p{\dimexpr 0.22\linewidth-2\tabcolsep\relax}>{\raggedright\arraybackslash}p{\dimexpr 0.22\linewidth-2\tabcolsep\relax}>{\raggedright\arraybackslash}p{\dimexpr 0.24\linewidth-2\tabcolsep\relax}>{\raggedright\arraybackslash}p{\dimexpr 0.19\linewidth-2\tabcolsep\relax}}
\caption{Immediate effect at the first divergence, executed on a copy of the pre-call state. The denominator for every fraction is the number of diverging trajectories in that row; 0/0 means no divergence and no applicable immediate-effect trials. New goal means that the native attack check becomes true after that single call; it is not a full counterfactual continuation.}\label{tab:app-authority-immediate-effects}\\
\toprule
\textbf{Variant} & \textbf{No tool error} & \textbf{State changed} & \textbf{New attack goal} & \textbf{Distinct tasks} \\
\midrule
\endfirsthead
\multicolumn{5}{l}{\small\tablename~\thetable{} (continued)}\\
\toprule
\textbf{Variant} & \textbf{No tool error} & \textbf{State changed} & \textbf{New attack goal} & \textbf{Distinct tasks} \\
\midrule
\endhead
\midrule
\multicolumn{5}{r}{\small Continued on next page}\\
\endfoot
\bottomrule
\endlastfoot
V1 & 0/0 & 0/0 & 0/0 & 0 \\
V2 & 0/0 & 0/0 & 0/0 & 0 \\
V3 & 0/0 & 0/0 & 0/0 & 0 \\
V4 & 110/110 & 110/110 & 4/110 & 20 \\
\end{longtable}
\endgroup

There is a different denominator for executing every statically reconsidered denial. Under V4, independently executing all 241 denied calls on their original pre-call states yields 5 new native attacker goals and 3 tool errors. First-divergence replay instead executes only the first changed decision in each trajectory: 110 calls, 4 new goals, and 0 tool errors. A later denied call can therefore appear in static re-evaluation but not first-divergence replay; these totals answer different questions and are not contradictory. Neither experiment estimates final task utility after a model responds to the changed decision.

Zero decision changes under V1--V3 show that the recorded denied writes do not match any configured format, city, or ASCII identity normalization rule. The single Eqv-classified denied call concerns an identity field that the native tool does not read; none of V1--V3 admits it. An identity probe also shows that case folding can equate a distinct recipient, so these zero changes do not establish that the relaxed rules are safe.

The verifier independently reconstructs classes, denominators, and first divergences from recorded evidence; its stored report contains 18 checks. For 86 allowed calls the state digest is clock-unverifiable and is not counted as a match. For 115 denied calls the pre-call state contains clock fields, which are masked for immediate-effect comparison; participant-item ordering is also normalized as described in the protocol. These qualifications do not establish byte-identical state for every call. Equivalence checks support bounded claims about the native task environment, not a claim of human-authorized equivalence or production safety. Eqv, Unres, and NoRef cases remain subject to author interpretation of the underlying task intent.